\documentclass[onecolumn]{aastex6}

\begin{document}

\title{Curvature induced polarization and spectral index behavior for PKS 1502+106}



%

\author{Xi Shao\altaffilmark{1}, Yunguo Jiang\altaffilmark{1} and Xu Chen\altaffilmark{1}}
\affil{Shandong Provincial Key Laboratory of Optical Astronomy and Solar-Terrestrial Environment,\\
Institute of Space Sciences, Shandong University, Weihai, 264209, China; jiangyg@sdu.edu.cn}
\begin{abstract}

A comprehensive study of multifrequency correlations  can shed light on the nature of variation for blazars. In this work, we collect the long-term radio, optical and $\gamma$-ray light curves of PKS 1502+106. After performing the localized cross-correlation function analysis, we find that correlations between radio and $\gamma$-ray or $V$ band are beyond the $3\sigma$ significance level. The lag of the $\gamma$-ray relative to 15 GHz is $-60^{+5}_{-10}$ days, translating to a distance $3.18^{+0.50}_{-0.27}$ parsec (pc) between them. Within uncertainties, the locations of the $\gamma$-ray and optical emitting regions are roughly the same, and are away from the jet base  within $1.2$ pc. The derived  magnetic field in optical and $\gamma$-ray emitting regions is about $0.36$ G. The logarithm of $\gamma$-ray flux is significantly linearly correlated with that of $V$ band fluxes, which can be explained by the synchrotron self-Compton (SSC) process, the external Compton (EC) processes, or the combination of them. We find a significant linear correlation in the plot of $\log\prod$ (polarization degree) versus $\log \nu F_{\nu}$ at $V$ band, and use the empirical relation $\Pi \sim \sin^n \theta'$ ($\theta'$ is the observing angle in the comoving frame blob) to explain it.
The behaviors of color index (generally redder when brighter at the active state) and $\gamma$-ray spectral index (softer when brighter) could be well explained by the twisted jet model.  These findings suggest that the curvature effect (mainly due to the change of the viewing angle) is dominant in the variation phenomena of fluxes, spectral indices, and polarization degrees for PKS 1502+106.

\end{abstract}

\keywords{galaxies: quasars: individual (PKS 1502+106) --- galaxies: jets --- $\gamma$-rays: general --- polarization: general}



\section{Introduction} \label{sec:intro}

Blazars are active galactic nuclei  (AGNs) with emitting jets pointing to our line of sight, which result in relativistic beaming of radiation \citep{1995pasp...107...803U}. Blazars are famous for their high luminosity, rapid variation, and high polarization. Their broadband  spectral energy distributions (SEDs) indicate radiation from radio to $\gamma$-ray, characterized by two prominent bumps. It is widely accepted that the first bump is caused by synchrotron radiation, while the second one is due to the inversely Compton scattered photons either from the synchrotron radiation in the jet or from external sources outside the jet. However, the locations of emitting regions for these two bumps have been under intensive debate. The correlation analysis between radio and $\gamma$-ray helps shed light on the location of $\gamma$-ray emitting regions, based on a series of blazar monitoring programs \citep{Cohen:2014,FUHRMANN:2014,Max:2014a}. However, the variation mechanisms to elucidate the color index and polarization behaviors are still away from convincing.

PKS 1502+106 (historically OR 103, S3 1502+10 and 4C 10.39) is a single-sided, core-dominated radio-loud AGN, and is classified as a flat spectrum radio quasar, located at R.A. = 15:04:25.0, decl. = +10:29:39. Its redshift is 1.839 \citep{Smith:1977,1980ApJ...235...361R}. The target underwent a strong high energy outburst in 2008 August, which was reported in ATel \#1650, followed by  multiwavelength campaigns at radio, visual, ultraviolet, and X-ray bands. The study of the target using Very Long Baseline Interferometry (VLBI) observations indicated the misalignment of radio components \citep{2004aa...421...839A}  and  jet bending morphology \citep{Karamanavis2016}. For this target, the multiple radio bands analysis is used to determine the opacity structure, and helps localize the $\gamma$-ray emitting region \citep{ Max:2014a,FUHRMANN:2014,Karamanavis2016}.  \citet{2014ApJs...215...5K} reported that $\gamma$-ray emitting regions are most likely beyond the broad line region (BLR) by studying a low frequency peaked blazar sample. They found that SED with seed photons from dust torus is better fitted than those from BLR. However, the SED fitting is not unique in principle. The optical emitting regions cannot be obtained directly from images, since almost all blazars are point sources. The correlation analysis between multiple frequencies may require the use of long-term time series to overcome the difficulty of low spatial resolution.

The intense variation study offers essential insights not only into the emitting regions but also into the intrinsic radiation processes. \citet{Abdo:2010} investigated radiation from radio to $\gamma$-ray of PKS 1502+106, and found that synchrotron and self-synchrotron Compton (SSC) processes dominate from radio to X-ray in SED, while $\gamma$-ray radiation is caused by the external Compton (EC) process.
The color index behavior in variation is another aspect to reveal the emission mechanism.  \citet{2006aa...453...817V} and \citet{2010pasj...62...645S} stated that the accretion disk emission contributing to the observed flux leads to the bluer when brighter (BWB) trend in outburst state and the redder when brighter (RWB) trend in active state for 3C 454.3. The similar behavior of CTA 102 studied by \citet{2017nature...552...374R} also can be  explained by the same theory. The variation of polarization is also important and can help us to further constrain the radiation models. In this work, we find a significant correlation between polarization degree (PD)  and fluxes for the target. Combining with correlations of multiband light curves, the color index, and $\gamma$-ray spectral index behaviors, we propose that the geometrical curvature effect can lead to various variation phenomena for the target in an unified manner.

This paper is organized as follows. In Section \ref{sec:data}, the $\gamma$-ray, optical $V$ and $R$ band, radio 15 GHz and PD data with periods of nearly nine years are collected. The optical data are calibrated using one meter telescope in the Weihai Observatory (WO).  The localized cross-correlation function (LCCF) among different time series are calculated, and time lags between them are derived. In Section \ref{sec:local}, we obtain the localization of the $\gamma$-ray and optical emitting regions, and  discuss their positions relative to the BLR.  In Section \ref{sec:discussion}, various variation phenomena and their correlations are discussed. Finally, our conclusion is given in Section \ref{sec:conclusion}.

%

\section{Data Reduction and Analysis}{\label{sec:data}}

\subsection{Data Reduction}
The {\it Fermi} Large Area Telescope (LAT) is a highly sensitive instrument with large viewing fields. Its survey scanning mode views the whole sky every 3 hr. The scanning cadence makes it an ideal  piece of equipment  to monitor $\gamma$-ray sources. We collect nearly nine years of $\gamma$-ray data (from MJD 54684 to MJD 57932) with energy range  $0.1-300$ GeV from the LAT data server.\footnote{\url{{https://fermi.gsfc.nasa.gov/}}.} The data was reduced using Fermi Science Tools version v10r0p5, and was analyzed by adopting the unbinned likelihood method. A $15^{\circ}$ region of interest (ROI) centered on the PKS 1502+106 was considered. To count the $\gamma$-ray background calculation, the galactic diffuse emission model (\rm{gll\_iem\_v06.fits}) as well as the extragalactic isotropic diffuse emission model (iso\_P8R2\_SOURCE\_V6\_v06.txt) were applied in the likelihood analysis. Furthermore, {\it make3FGLxml.py} was used to create a source model file.  The instrument response function was chosen to be P8R2\_SOURCE\_V6. The flux information of the target under the filtering condition $\rm{TS}>10$ were extracted from the {\it gtlike} result files. In addition, the $\gamma$-ray spectral index in one time bin was obtained by linearly fitting fluxes of seven energy bins, which are logarithmically divided in the range of $0.1-218.7$ GeV. This will minimize the correlation between flux and its index in the likelihood analysis.\par

\begin{figure}
  \centering
 \includegraphics[scale=0.1]{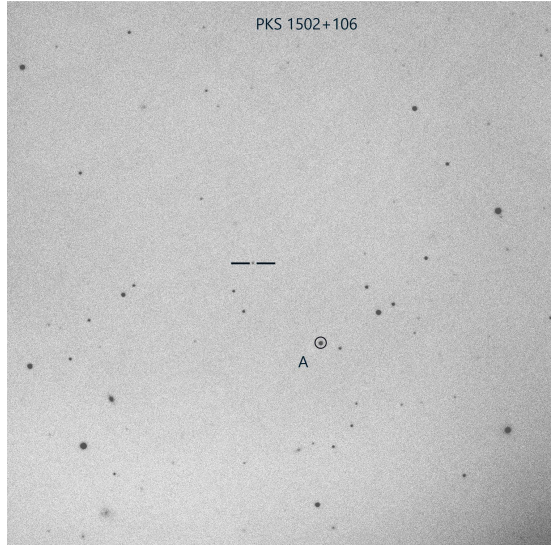}
 \caption{$R$ band image of PKS1502+106 (centered between two bars) and its comparison star A (in circle) is obtained from the one-meter telescope of the Weihai Observatory.}\label{Fig:FC}
\end{figure}

\begin{figure}[htbp]
 \centering
 \includegraphics[scale=0.7]{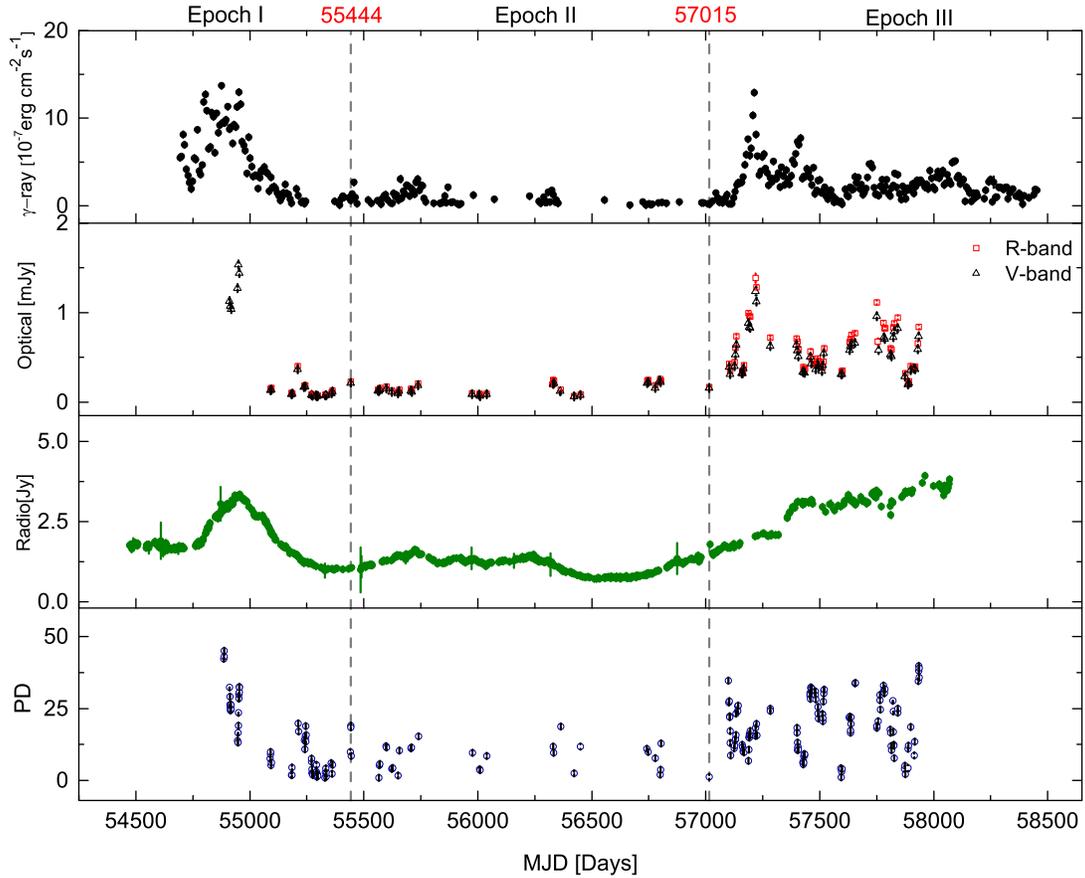}
 \caption{From top to bottom, the light curves of $\gamma$-ray, optical $V$ band and $R$ band, radio 15GHz, and PD are plotted, respectively. The two vertical dashed lines (MJD 55444 and 57015) divide light curves into three periods, namely Epoches
  I, II and III, which correspond to the giant flare, the quiescent, and the active state, respectively.  }\label{Fig:LC}
\end{figure}

{Steward Observatory (SO) has monitored this target for a long period since 2009 February 24 (MJD 54520), which  provides the optical $V$ band, $R$ band, and polarization data.\footnote{\url{http://james.as.arizona.edu/~psmith/Fermi}} However, as for the photometry data, only the differential magnitudes in $V$ and $R$ bands of the target are available, and the comparison star A in the finding chart was not calibrated \citep{Smith:2009}. Using the one-meter Cassegrain telescope at the WO, we performed the photometric observations for the comparison star A and the other six Landolt standard stars on 2019 January 16 (MJD 58499). The telescope was mounted with the Johnson/Cousins set of {\it UBVRI} filters and the back-illuminated PIXIS 2048B CCD camera with  2k$\times$2k square pixels \citep{Hu:2014}. The field of view is about $11'.8\times11'.8$. The images  obtained were corrected with the bias and flat field, and  the $R$ band finding chart image for the comparison star A and PKS 1502+106 is presented in Figure \ref{Fig:FC}. Following the standard aperture photometric procedure using {\verb"IRAF"},\footnote{IRAF is distributed by the National Optical Astronomy Observatories, which is operated by the Association of Universities for Research in Astronomy Inc., under contract to the National Science Foundation} the apparent magnitudes of the comparison star A  were calculated to be $B=16.493\pm0.029$, ${V}=15.521\pm0.017$, and $R=14.984\pm0.022$, respectively.  $U$ band data was abandoned due to low signal-to-noise ratio, while the apparent magnitude in $I$ band was not considered due to its less significant fitting of extinction coefficients. In addition, the galactic extinctions in $V$ and $R$ bands are 0.106 and 0.086 magnitude, respectively.\footnote{\url{http://ned.ipac.caltech.edu/}} The fluxes of targets are calculated based on the differential photometry data and the calibration results above. The flux errors are inherent only from the differential photometry measurement errors. The light curves obtained  are shown in Figure \ref{Fig:LC}. The SO used the SPOL to derive the Stokes parameters ($q=Q/I$ and $u=U/I$). The fractional $q$ and $u$ have been calibrated without considering the interstellar polarization. In this work, we consider only the light curve of polarization degree (PD). The polarization angle (PA) has the $n\pi$ ambiguity problem \citep{Marscher:2008,Kiehlmann:2016}, and the large gap between observations will reduce its validity of variation especially for the long period.  }

 As for radio data, we collect the calibrated data from the OVRO 40 m monitoring program \citep{Richards:2011}. The data period are  from 2008 January 8 (MJD 54473) to 2017 November 12 (MJD 58099) with 630 points. The long-term duration will enhance the significance of the correlation analysis. The light curves of $\gamma$-ray, optical, radio, and PD are shown in Figure \ref{Fig:LC}.

\subsection{Data Analysis}
\begin{figure}[ht]
    \begin{minipage}[t]{0.48\linewidth}
        \centerline{\includegraphics[scale=0.35]{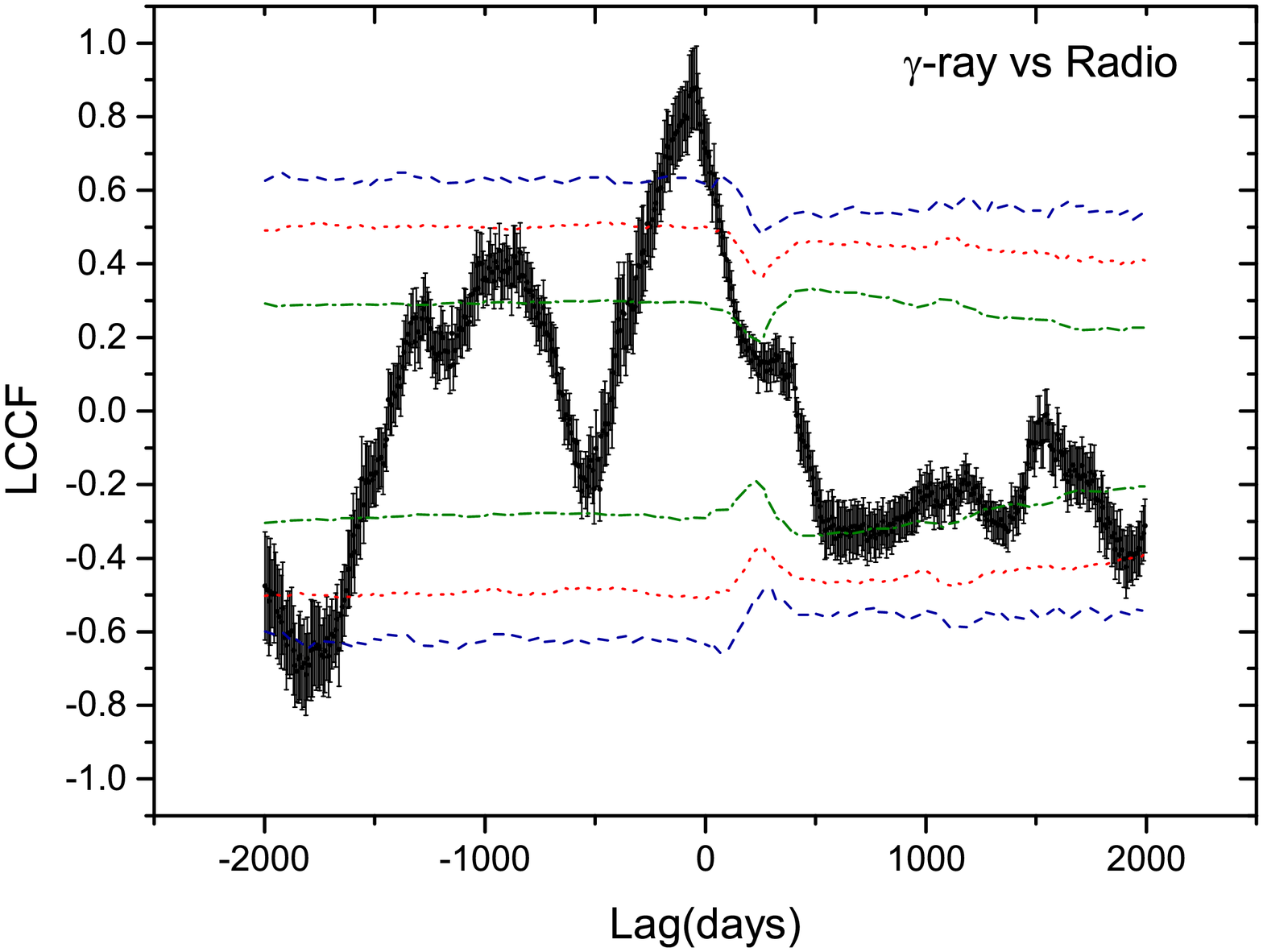}}
           \end{minipage}%
    \begin{minipage}[t]{0.496\linewidth}
        \centerline{\includegraphics[scale=0.35]{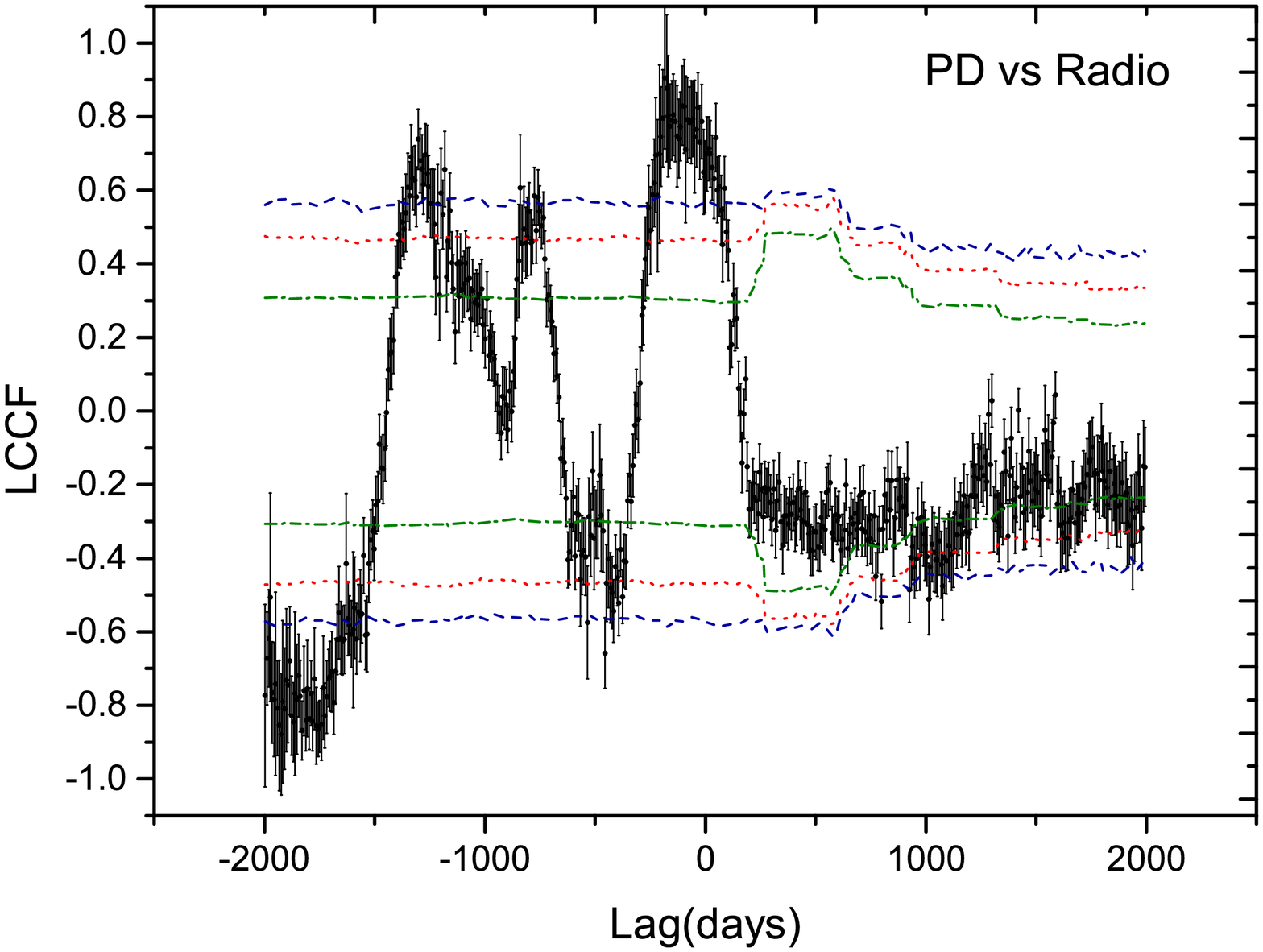}}
    \end{minipage}%

    \begin{minipage}[t]{0.48\linewidth}
        \centerline{\includegraphics[scale=0.35]{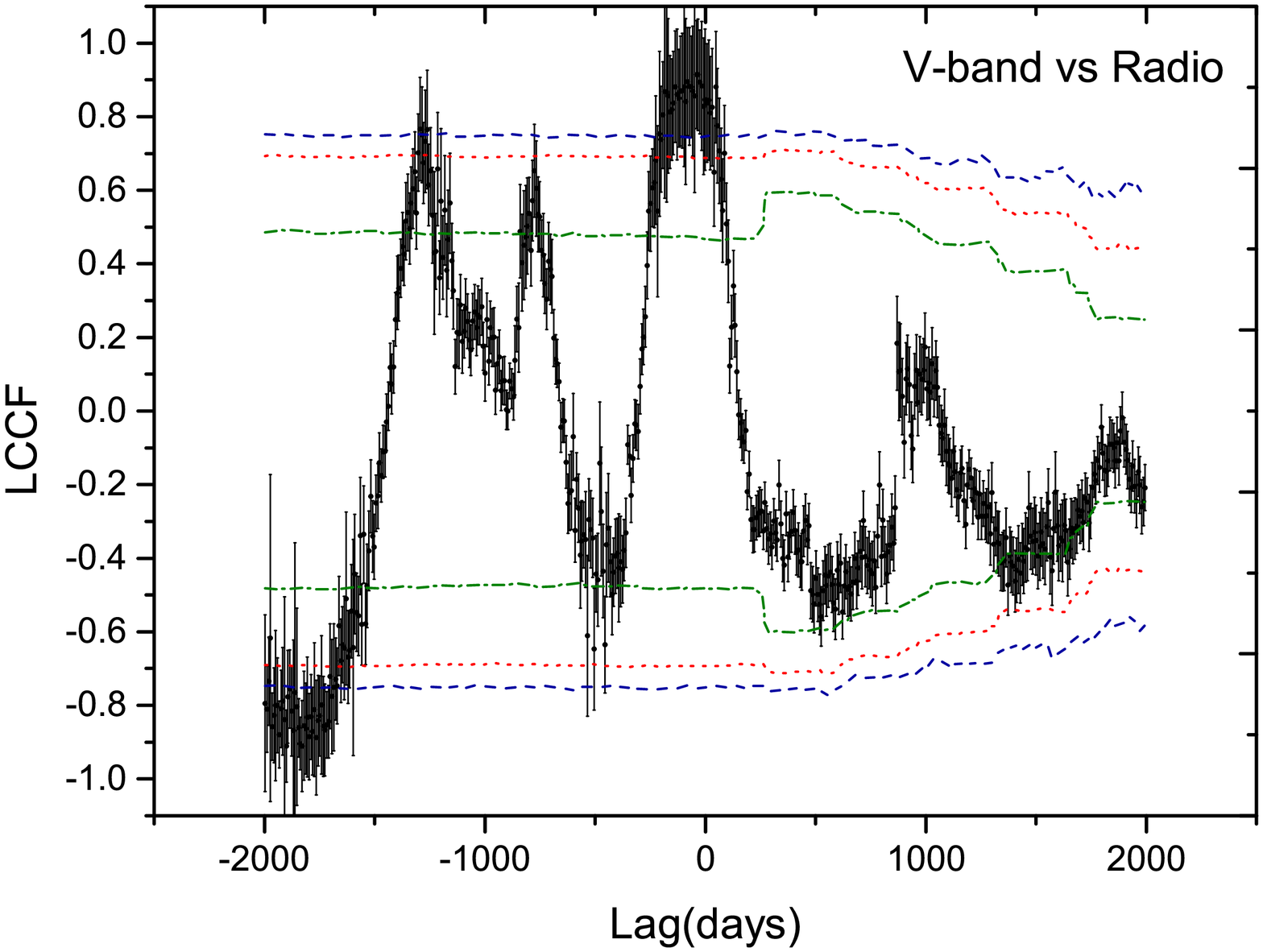}}
    \end{minipage}
    \begin{minipage}[t]{0.48\linewidth}
        \centerline{\includegraphics[scale=0.35]{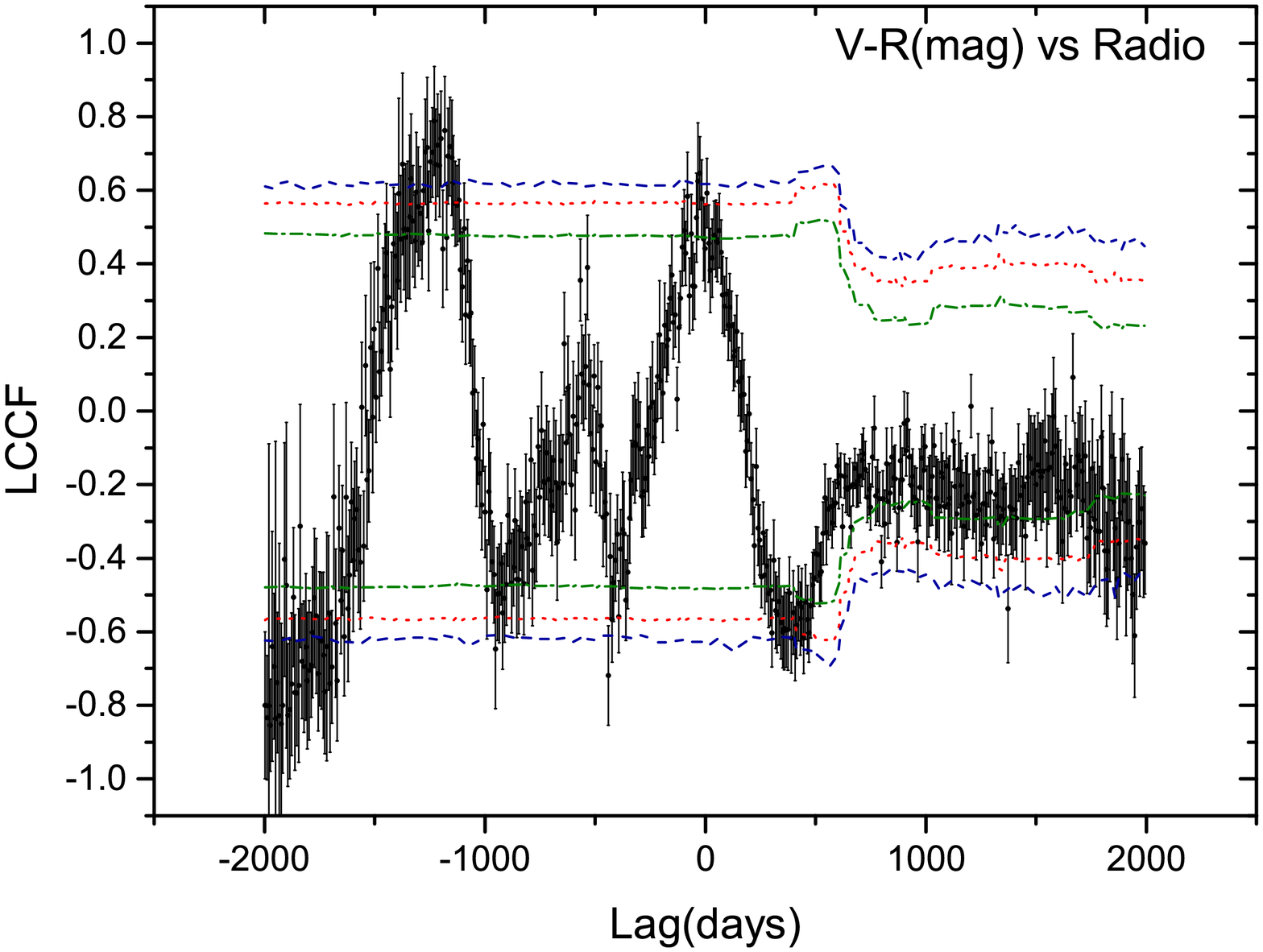}}
    \end{minipage}
    \caption{Upper left and upper right panels present the plots of LCCF results of $\gamma$-ray vs. radio (15 GHz) and PD versus radio, respectively. The lower left and right panels are LCCFs between $V$ band flux and radio and $V-R$ (magnitudes) vs. radio, respectively. Black dots with error bars donate correlation values. The olive dashed dotted, red dotted and royal blue dashed lines donate the $1\sigma$, $2\sigma$, $3\sigma$ significance levels, respectively. The negative lag indicates that the former leads to the latter. }\label{Fig:LCCF}
\end{figure}

One popular approach to calculate the correlation between unevenly sampled light curves is the discrete correlation function (DCF; \cite{Edelson:1988}). The absolute value of DCF can be larger than unity. Another normalized correlation function is invented by \citet{Welsh:1999}, which is named as the LCCF, and is given by
\begin{equation}
{\rm LCCF}(\tau)=\frac{1}{M}\sum_{ij}\frac{(a_i-\bar{a}_{\tau})(b_j-\bar{b}_{\tau})}{\sigma_{a\tau} \sigma_{b\tau}},
\end{equation}
where $a_i$ and $b_j$ are time series, $M$ denotes the number of ($a_i$, $b_j$) pairs which satisfy the condition $\tau \leq \Delta t_{ij}\leq \tau +\delta t$ ($\delta t$ is the bin time), $\bar{a}_{\tau}$ and $\bar{b}_{\tau}$ are the averaged values of the M pair samples, and $\sigma_{a\tau}$ and $\sigma_{b\tau}$ are the corresponding standard deviations. Following the definition of DCF uncertainty, the error of the LCCF coefficient is taken to be the standard deviation of the local M samples, i.e.,
\begin{equation}\label{eq:error}
\sigma_{\rm LCCF}(\tau)=\frac{1}{M-1}\left( \sum[{\rm LCCF}_{ij}-{\rm LCCF}(\tau)]^2 \right)^{\frac{1}{2}}.
\end{equation}
The values of LCCF are in the range $[-1,1]$, which is a good property to be used in significance estimation. Compared with DCF, the LCCF is more efficient to pick up physical signals \citep{Max:2014b}.
Thus, LCCF will be used to analyze correlations between various time series of PKS 1502+106. The sampling of the optical observation is extremely uneven, while radio observation has relatively even samplings. Thus, we bin the optical and radio light curves with $4$ and $7$ day intervals, respectively. The $7$ day bin for  the optical curve has also been tested, and no obvious change in the LCCF result is evident. In the reduction procedure, a time step of 7 day has already been considered to produce the $\gamma$-ray light curve. The light curve of PD is not binned as that of fluxes, since the binning process for PD is of nonlinear and could lead to spurious correlations. The rebinning procedure can smooth the LCCF profile and the significance levels. Although interpolation can reduce the red-noise leakage problem, it can also bring spurious signals especially when there are large observation gaps. Thus, no interpolation has been applied to all observed data.
The LCCFs of $\gamma$-ray, optical, PD, $V-R$ (magnitude) versus radio  are plotted in Figure \ref{Fig:LCCF}. The range of lag time is taken to be [$-2000,2000$] with an 8 day bin.

The Monte Carlo (MC) simulation to produce the significance levels is essential to interpret the cross-correlation result.
In our recipe, we simulate 10,000 artificial light curves using the TK 95 algorithm with $\beta=2.3$  \citep{Timmer:1995}, which is an ensemble of the radio light curve.
Each light curve contains 3000 points separated by equal bins of 2 days. Then LCCFs between artificial light curves and the observed one are calculated to produce a distribution at each lag bin. The $1\sigma$ (68.27\%), $2\sigma$ (95.45\%), and $3\sigma$ (99.73\%) significance levels are marked with olive dashsed dotted, red dotted and royal blue dashed lines in Figure \ref{Fig:LCCF}, respectively. This step is different from that of \citet{Cohen:2014} and \citet{Max:2014a}, who used the completely observed and completely artificial pairs, respectively. The aim of significance estimation is to find the chance probability for a correlated physical signal in a random sample. In some sense, our procedure can reduce the impact of sampling affects from observation, and can escape from the red-noise leakage problem in PSD for optical and $\gamma$-ray data.

\citet{Max:2014a} obtained that the $3\sigma$ significance range in the $\gamma$-ray versus radio MC approaches $0.9$, which is larger than our result.  They concluded that the peak of correlation coefficients is below the $3\sigma$ level.  First, the main difference between our results and \citet{Max:2014a} stems from the assumed $\beta_{\gamma}=1.6$ in their simulation. \citet{Abdo:2010} analyzed the PSD of the $\gamma$-ray light curve with a period of 140 days for PKS 1502+106, and obtained  $\beta_{\gamma}=1.3$. A flatter PSD will decrease LCCF coefficients, and further reduce the significance range. This reason is evident in the comparison between PKS 1502+106 and other two sources (AO 0235+1164 and B2 2308+34). Second, our LCCF peak is a little higher than \citet{Max:2014a}, since we use nearly 9 yr of data to perform LCCF. The long duration of the observation will enhance coefficients of correlation, when two curves have physical coherence. The TK95 algorithm considers only the PSD to produce the artificial light curve, whose fluxes have a Gaussian distribution. However, the fluxes of observed light curves usually are non-Gaussian distributed. Such property is characterized by the probability density function (PDF), which can also affect the confidence level. \citet{Emmanoulopoulos:2013} indicated that the significance for the peak of DCCF is more conservative, if both PSD and PDF are considered in the simulation. Thus, significance levels in our recipe are still underestimated to some extent.

\begin{figure}[ht]
    \begin{minipage}[t]{0.48\linewidth}
        \centerline{\includegraphics[scale=0.35]{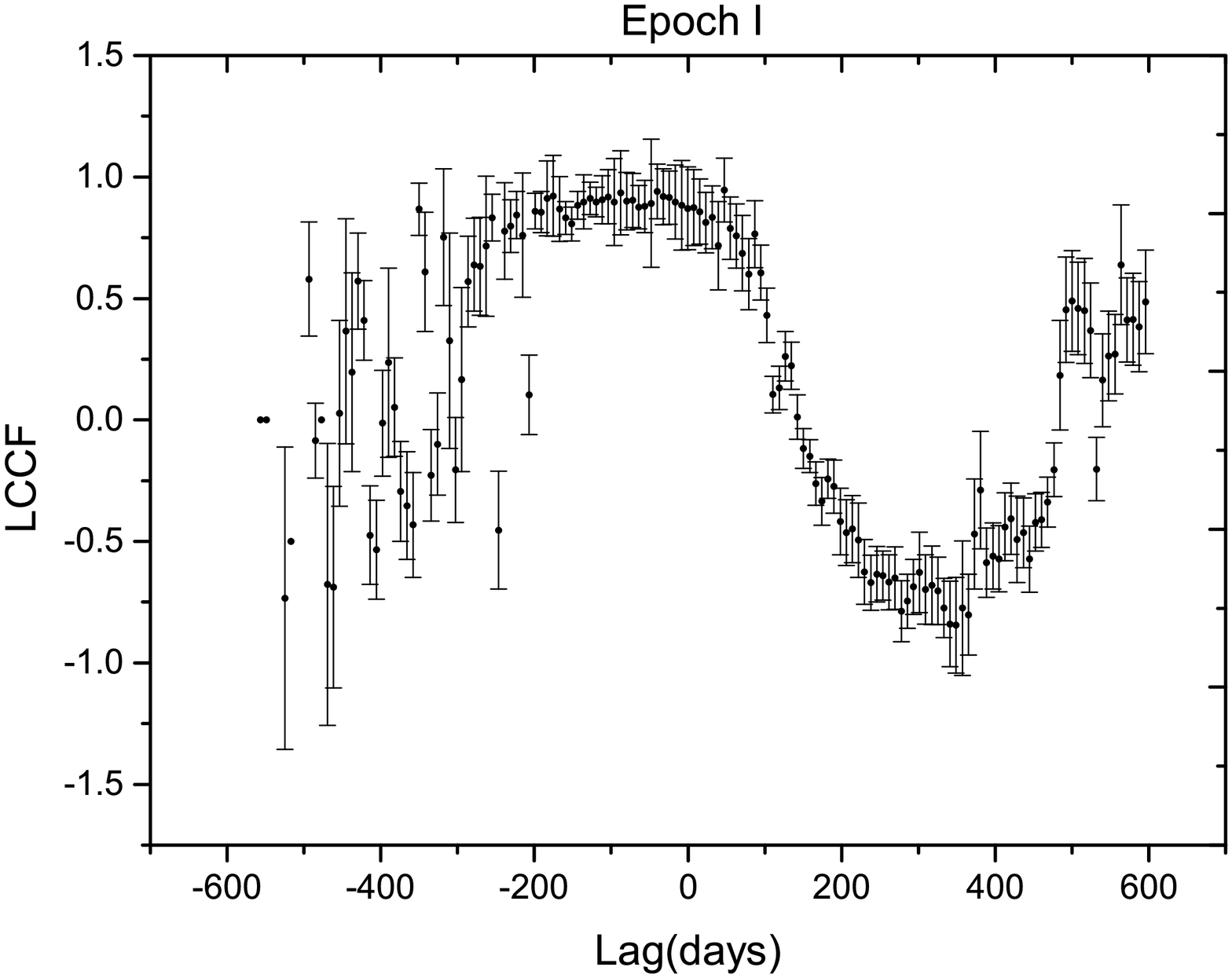}}
    \end{minipage}%
    \begin{minipage}[t]{0.48\linewidth}
        \centerline{\includegraphics[scale=0.35]{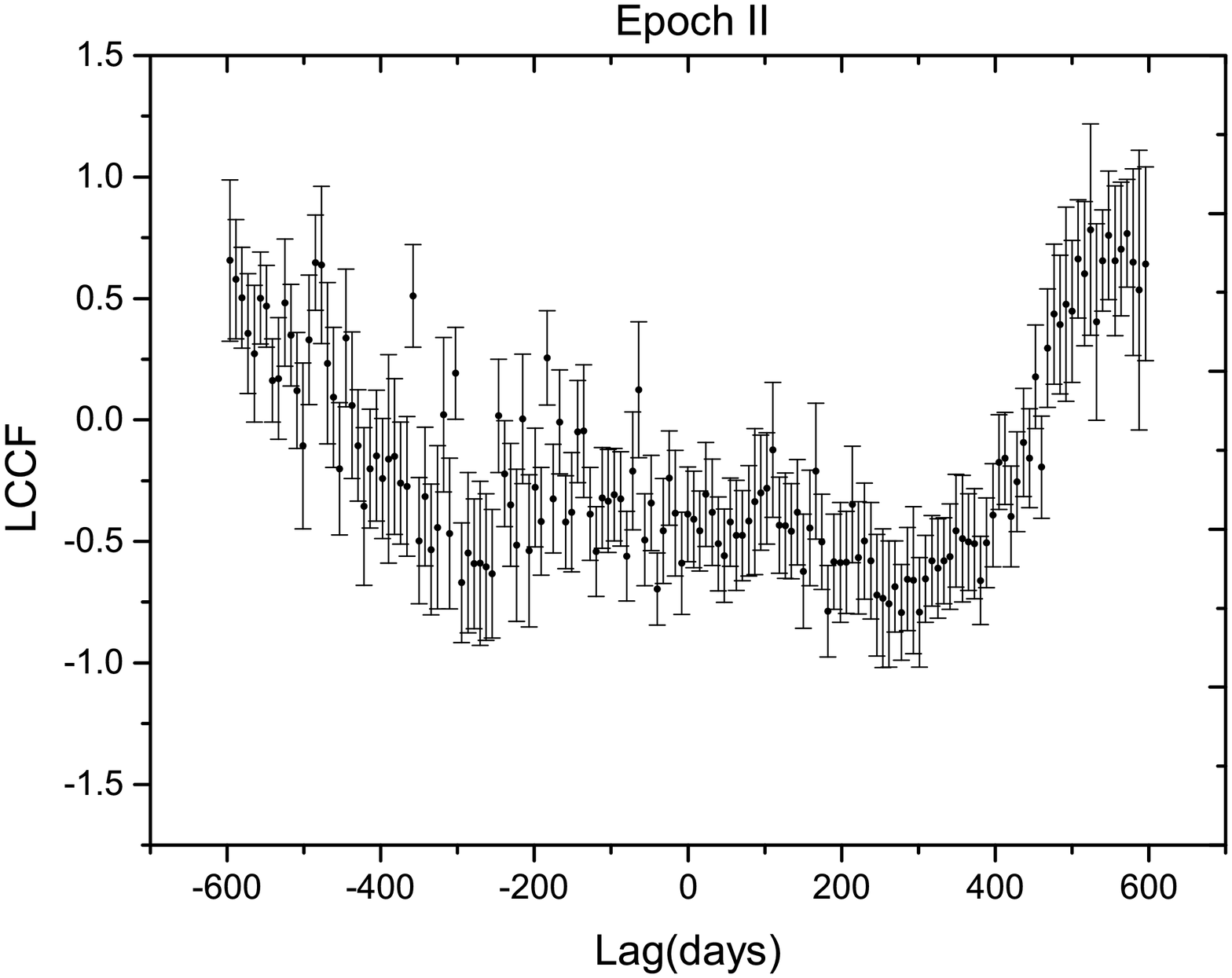}}
    \end{minipage}%
    \caption{Left panel is the plot of LCCF of $V$ band flux vs. radio data in Epoch I, and the right panel is that in Epoch II.}\label{Fig:LCCFVR_III}
\end{figure}

\begin{figure}[tp]
  \centering
  \includegraphics[scale=0.35]{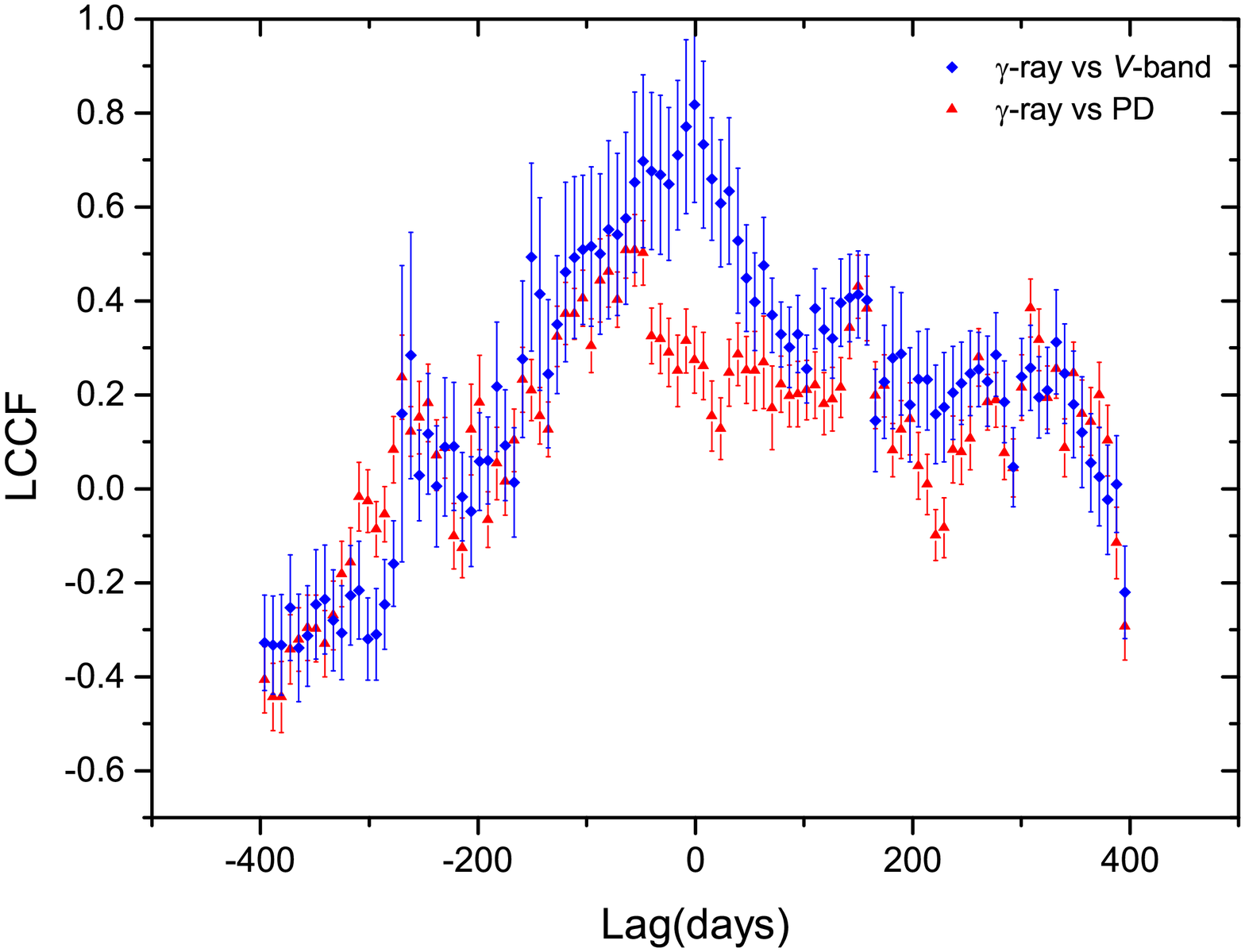}
  \caption{LCCFs of $\gamma$-ray vs. $V$ band flux (blue diamond) and $\gamma$-ray vs. PD (red triangle) are plotted in the time interval $[-400,400]$ with 8 day bin.  }\label{Fig:GammaV}
\end{figure}

In the upper left panel of Figure \ref{Fig:LCCF}, the sharp peak with beyond $3\sigma$ significance indicates the strong correlation between the $\gamma$-ray and radio light curves. And the variation of $\gamma$-ray leads a variation of radio of about dozens of days. In both the upper right panel and lower left panel, there are two peaks beyond the $3\sigma$ level, and the most significant peak shows a plateau, which spans less than 300 days. To elucidate the appearance of the plateau, we calculate LCCFs between optical and radio in Epochs I and II, respectively. The plateau occurs in Epoch I, and disappears in Epoch II, see Figure \ref{Fig:LCCFVR_III}.  We also plot LCCFs of $\gamma$-ray versus $V$ band (blue diamond) and $\gamma$-ray versus PD (red triangle) in Figure \ref{Fig:GammaV}. No significant lag is found between $\gamma$-ray and $V$ band flux, while $\gamma$-ray leads PD for about $50$ days, but the peak coefficient of LCCF (about $0.5$) is less significant.  In this plot, there is no plateau.  Notice that there are several complete flares in both the $\gamma$-ray and $V$ band light curves in Epoch III. Hence, the appearance of the plateau is most probably due to the missing rising phase of the giant flare at $V$ band in epoch I, because the correlation between a monotonically decreasing curve and a complete flare curve is invariant in the shift of lag time. The plateau brings trouble when the lag time is estimated. It is most likely that variations of $V$ band and $\gamma$-ray are simultaneous, and they both lead to variation of radio. Variation of PD is significantly correlated with variation of radio flux, which helps us to understand the variation mechanism. Detailed lag times will be analyzed in the following.

In the lower right panel of Figure \ref{Fig:LCCF}, the most significant peak (beyond $3\sigma$) of LCCF is located at $-1230$, while the second significant peak (about $2\sigma$) is located around zero. Connected with the significance analysis above, the peak around $-1230$ is a spurious signal. It is impossible that the variation color index leads that of the flux for nearly a thousand days. The magnitude of variation for the color index has no linear relation with that for flux generally. It is shown that $V-R$ has larger magnitude also in quiescent state in Figure \ref{figure6}, leading to a curved shape of color index light curve in Epoch II. The mismatched flares in two time series will lead to a spurious signal in LCCF. Even so, the second peak tells us that the color index varies in a similar cadence with the flux. For this target, the sparse samplings for optical observation hinder us from performing a time delay analysis between them.

The time lag estimation, together with its $1\sigma$ error range, is based on the  model-independent Monte Carlo method proposed by \citet{Peterson:1998}. The procedure considers the flux randomization (FR) and random subset selection (RSS; \cite{Peterson:1998,Lasson:2012}). Two kinds of time delays are considered, i.e., $\tau_p$ and $\tau_{c}$. $\tau_p$ is defined as  the lag corresponding to the peak of the LCCF.  $\tau_c$ is  the centroid lag of the LCCF defined as $\tau_c=\sum\limits_{i} \tau_i C_i/\sum\limits_{i} C_i$, where $C_i$ is the correlation coefficient satisfying  $C_i >0.8 {\rm LCCF}(\tau_p)$. We repeat $10^4$ times to obtain a distribution for $\tau_p$ and $\tau_{c}$. To better locate time delays, we set that the lag range of LCCF to [$-600,600$] and the lag bin to 4 days.  The errors of  time delays are of the $1\sigma$ range from the distribution \citep{Jiang:2018}.

The time delays $\tau_p$ and $\tau_c$ between different energy bands, together with their average $\langle \tau \rangle$, are shown in Table \ref{tab:lag}. The most significant time delay is between $\gamma$-ray and radio, i.e., $\tau_p=-40_{-8}^{+0}$ or $\tau_c=-80_{-11}^{+10}$. \citet{Abdo:2010} derived that peak flux of $\gamma$-ray arrives 98 days before that of  radio 15 GHz for the giant flare in Epoch I, which is close to our result $\tau_c=-80_{-11}^{+10}$. \citet{Max:2014a} obtained $\tau_p=-40\pm 13$ for $\gamma$-ray versus radio, which is almost the same as our result $\tau_p=-40_{-8}^{+0}$.

We derive $\tau_p=-72_{-103}^{+119}$ for $V$ band versus radio, which has large uncertainties due to the plateau. \citet{Karamanavis2016} measured the time lags among different radio frequencies using both DCF and Gaussian process regression (GP) for the target. The time delays between $15$ and $142.33$ GHz are $\tau_{\rm GP}=64.2$ and $\tau_{\rm DCF}=63_{-51}^{+43}$, respectively. This value is close to our result $\tau_c=-64_{-6}^{+7}$ for optical versus radio.  So the radiation of $142.33$ GHz and optical band may originate from the same optically thin regime.  We obtain that optical $V$ band lags behind $\gamma$-ray for $\tau_p=8_{-43}^{+12}$ or $\tau_c=17_{-16}^{+61}$. \citet{Abdo:2010} also obtained that $\gamma$-ray leads optical with 4 days using $\tau_p$ of DCF, which is roughly in accordance with our result. Within uncertainties, we conclude that $\gamma$-ray and optical photons  are most probably emitted from the same region.

 \begin{deluxetable}{ccccr}[b!]
\tablecaption{Time delays\label{tab:lag}}
\tablecolumns{5}
\tablenum{1}
\tablewidth{0pt}
\tablehead{
\colhead{Time Delays} & \colhead{$V$ band versus Radio} & \colhead{$\gamma$-ray versus Radio} & \colhead{$\gamma$-ray versus $V$ Band} \\
}
\startdata
  $\tau_p$ & $-72_{-103}^{+119}$ & $-40_{-8}^{+0}$ & $-8_{-43}^{+12}$ \\
  $\tau_c$ & $-64_{-6}^{+7}$ & $-80_{-11}^{+10}$ & $-17_{-16}^{+61}$ \\
    $\langle \tau \rangle $ & $-68_{-55}^{+63}$ & $-60_{-10}^{+5}$ & $-13_{-30}^{+37}$ \\
\enddata
\tablecomments{ $\tau_p$ and $\tau_c$ are all in units of days, $\langle \tau \rangle $ is the mean of $\tau_p$ and $\tau_c$.  A negative lag indicates that the former leads the latter.}
\end{deluxetable}

\section{Locations of the Optical and $\gamma$-Ray Emitting Regions}{\label{sec:local}}

\begin{deluxetable}{ccCcc}[b!]
\tablecaption{Relative distances \label{tab:D}}
\tablecolumns{5}
\tablenum{2}
\tablewidth{0pt}
\tablehead{
\colhead{Distance} & \colhead{$V$ band versus Radio} & \colhead{$\gamma$-ray versus Radio} & \colhead{$\gamma$-ray versus $V$ Band}\\
}
\startdata
  $ D_p$ & $3.82_{-6.32}^{+5.47}$ & $2.12_{-0}^{+0.42}$ & $0.42_{-0.64}^{+2.23}$\\
  $ D_c$ & $3.34_{-0.37}^{+0.32}$ & $4.24_{-0.53}^{+0.58}$ & $0.90_{-3.23}^{+0.85}$\\
  $\langle  D \rangle$ & $3.58_{-3.35}^{+2.90}$ & $3.18_{-0.27}^{+0.50}$ & $0.66_{-1.94}^{+1.54}$\\
\enddata
\tablecomments{ $\langle  D \rangle$ is the average of $ D_p$ and $ D_c$ in units of parsec. The positive value indicates that the emitting region of the former is  in the upstream of the latter. }
\end{deluxetable}

Assuming a canonical jet, a perturbation propagates along the jet, emitting $\gamma$-ray (for instance) at the upstream and radio emission at the downstream. The observed time delay between $\gamma$-ray and radio depends on the red-shift $z$, the viewing angle $\theta$, the velocity of perturbation $\beta=v/c$, and the distance $ D$ between their emitting regions (see Figure 3 in \citep{Max:2014a}). Analytically, the distances between different emission regions at different energy bands in the rest frame of the quasar are derived as  \citep{Kudryavtseva:2011,Max:2014a}
\begin{equation}\label{eq1}
 D=\frac{\beta_{app}c T}{(1+z)\sin\theta},
\end{equation}
where $\beta_{\rm app}$ is the apparent velocity in the observer frame, $c$ is the light speed, $T$ is the time delays ($\tau_p$ or $\tau_c$) between different bands in observer frame, and $z$ is the redshift, and $\theta$ is the viewing angle between jet axis and observing line of sight. {  For PKS 1502+106, \citet{Hovatta:2009} obtained $\beta_{\rm app}=14.7$ using the fastest knot feature. \citet{Pushkarev:2009} derived $\theta=4^{\circ}.7$ by the VLBA observation. The target has a redshift $z=1.839$ \citep{Smith:1977,Abdo:2010}. The jet parameters are estimated by variation of radio fluxes and knot features in images, and can vary in a range for different knot observations.

Corresponding to $\tau_p$ and $\tau_c$, we define two kinds of relative distance $D_p$ and $D_c$ via Equation \ref{eq1}. The derived relative distances, i.e., $D_p$, $ D_c$ and their average $\langle  D \rangle $, are summarized in Table \ref{tab:D}. We obtain that the $\gamma$-ray emitting region is at upstream of the jet, separating from the radio emitting region with distance $3.18_{-0.27}^{+0.50}$ \,\rm{parsec} (pc) (corresponding to $\langle \tau \rangle$).
Meanwhile, relative distance between $\gamma$-ray and $V$ band is $0.66_{-1.94}^{+1.54}$ pc, which indicates that the emitting regions of optical and $\gamma$-ray are very close in jet.

To derive distances between emitting regions and the jet base, we need to refer to $r_{\rm core}$, which denotes the distance between the 15 GHz core region and the jet base.
\citet{2012aj...137...3718P} presented $r_{\rm core}=8.19$ pc for PKS 1502+106.  \citet{Karamanavis:2016} presented $r_{\rm core}=3.8\pm0.7 \rm{pc}$ for 15GHz emissions, using the time lag based core shift measurement, which combines the proper motion and time lags to derive core position offset. The mass of the black hole in PKS 1502+106 is about $10^9 { M}_{\odot}$, and the BLR region size is about $0.1$ pc. If one takes $r_{\rm core}=8.19$ pc, then the $\gamma$-ray emitting region is located about 5 pc away from the jet base, which is far away from the BLR region. If one takes
 $r_{\rm core}=3.8\pm0.7 \rm{pc}$, then the distance between the $\gamma$-ray (possibly and optical) emitting region and the jet base is less than $1.2$ pc. The possibility that $\gamma$-rays are emitted inside the BLR region cannot be ruled out.

 %
Additionally, the magnetic field of emitting zones is derived from the formula $B=B_1r^{-1}$, where $B_1$ is given by \citep{OSullivan:2009}
\begin{equation}\label{eq3}
B_1\simeq0.025\bigg(\frac{\Omega^3_{r\nu}(1+z)^2}{\varphi\delta^2\sin^2\theta}\bigg)^{1/4},
\end{equation}
which represents the magnetic filed at 1 pc away from the jet base. {  For this target, \citet{2012aj...137...3718P} presented $B_1=0.69\rm{G}$ via core shift measurement, while \citet{Karamanavis:2016} presented $B_1=0.18\rm{G}$ via time lag core position offset. Referring to parameters given by \citet{Karamanavis:2016}, the magnetic field in $\gamma$-ray and  optical emitting regions is about $0.36\rm{G}$.}

\section{Discussion of Variations}{\label{sec:discussion}}
\subsection{Optical $V$ Band and $\gamma$-Ray}
We aim to figure out the  {emission mechanism} at optical and $\gamma$-ray bands. {The observed fluxes of synchrotron, SSC and EC, which mainly depend on three parameters, i.e., the particle number density $N_e$, the magnetic field strength $B$ and the Doppler factor $\delta$, are given by}\footnote{To derive Equations (\ref{eq4})-(\ref{eq6}), the convention is that $F_{\nu}\propto \nu^{-\alpha}$.} \citet{2012aj...749...271C}
\begin{equation}\label{eq4}
  F_{syn}\sim N_eB^{1+\alpha_o}\delta^{3+\alpha_o},
\end{equation}
\begin{equation}\label{eq5}
   F_{EC}\sim N_e\delta^{4+2\alpha_{\gamma}}U'_{ext},
\end{equation}
\begin{equation}\label{eq6}
  F_{SSC}\sim N_e^2B^{1+\alpha_o}\delta^{3+\alpha_{\gamma}},
\end{equation}
where $\alpha_o$ is the spectral index at optical band, and $\alpha_{\gamma}$ is that of {$\gamma$-ray}, and $U'_{ext}$ is the energy density of external photons in the jet comoving frame. Taking the logarithm of these fluxes, three parameters can be disentangled. For instance, $\log F_{syn}= \log N_e + (1+\alpha_o)\log B+ (3+\alpha_o)\log \delta+C$, where $C$ is a constant. The optical $V$ band radiation  is of synchrotron, while $\gamma$-ray  photons are produced by SSC or EC in the lepton model. We select pairs of $\gamma$-ray and $V$ band fluxes observed on the same date (the time difference is less than 2 days), and plot $\log E^2 dN / dE$ ($\gamma$-ray) versus $\log\nu F_{\nu}$ ($V$ band) in Figure \ref{figure10}. If variation of $B$ dominates, one has $\Delta \log F_{syn} \sim (1+\alpha_o) \Delta \log B$, $\Delta \log F_{\rm EC} \sim 0$ and $\Delta \log F_{\rm SSC} \sim (1+\alpha_o) \Delta \log B$. Thus, one can expect that the slope in the plot is $1$ for the SSC process, and no linear correlation should be found for the EC process. If the Doppler factor is the dominant variable, then one can derive the slope for the SSC versus synchrotron case to be $\frac{3+\alpha_{\gamma}}{3+\alpha_o}$.  All other theoretical slopes are derived and summarized in Table \ref{Tab:index}.

 \begin{deluxetable}{ccCcc}
\tablecaption{Theoretical correlations \label{Tab:index}}
\tablecolumns{4}
\tablenum{3}
\tablewidth{0pt}
\tablehead{
\colhead{Cases} & \colhead{$ N_e$} & \colhead{$ B$} & \colhead{$\delta$}\\
}
\startdata
  $ F_{\rm SSC} \sim F_{\rm synch}$ & $2$ & $1$ & $\frac{3+\alpha_{\gamma}}{3+\alpha_o}$\\
  $ F_{\rm EC} \sim F_{\rm synch}$ & $1$ & $-$ & $\frac{4+2\alpha_{\gamma}}{3+\alpha_o}$\\
  $\Pi \sim F_{\rm synch}$ & $-$ & $-$ & $\frac{n(1+\alpha_{\theta})}{3+\alpha_o}$\\
\enddata
\tablecomments{ {  The predicted correlations in the plot of $\gamma$-ray versus optical fluxes in logarithm for different processes are plotted. {The symbol} '$-$' denotes that there should be no correlation. $\Pi$ is optical polarization degree, and derivations are referred to Section \ref{Sec:pol}. } }
\end{deluxetable}

 \begin{figure}[ht]
    \begin{minipage}[t]{0.48\linewidth}
        \centerline{\includegraphics[scale=0.35]{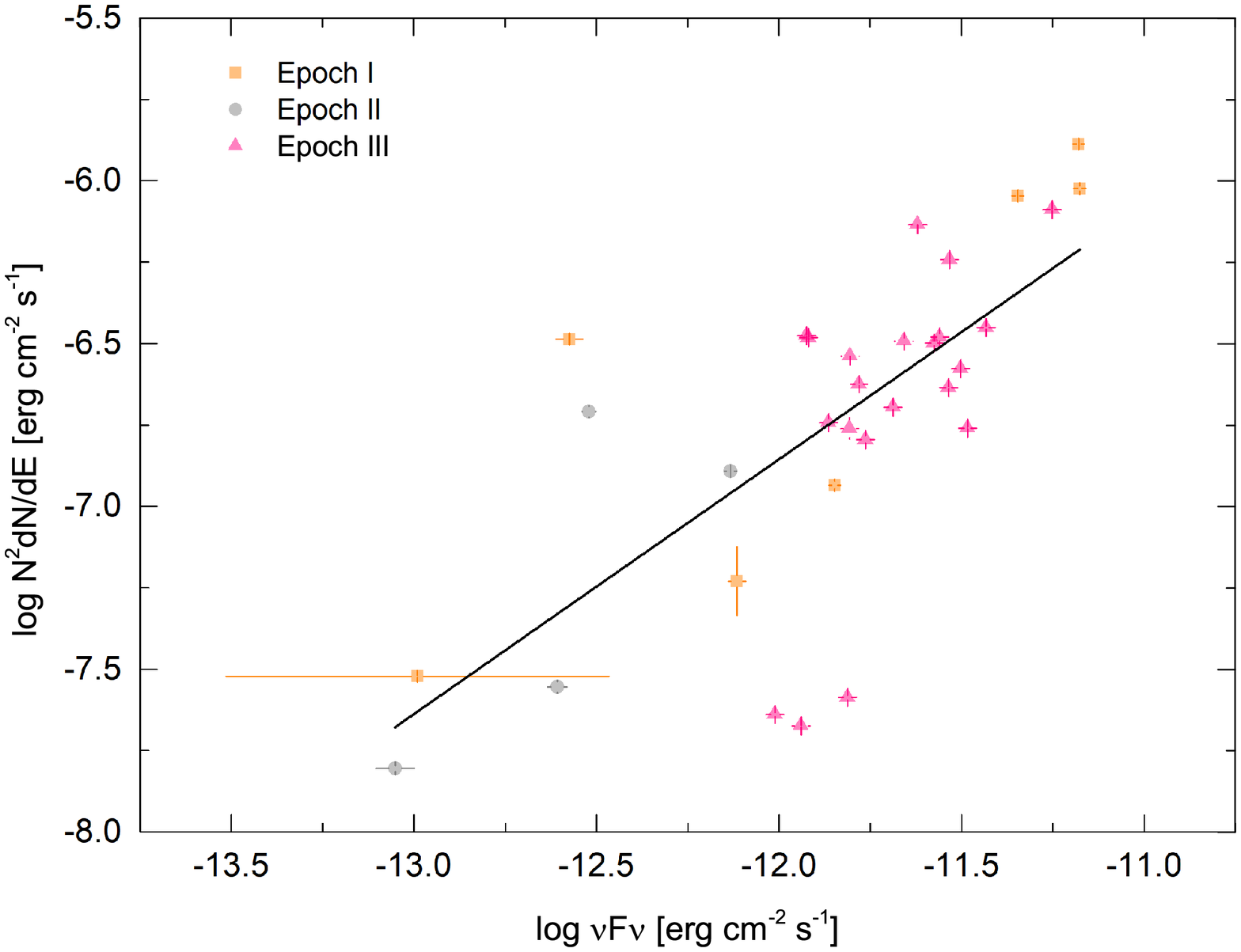}}
    \end{minipage}%
    \begin{minipage}[t]{0.48\linewidth}
        \centerline{\includegraphics[scale=0.35]{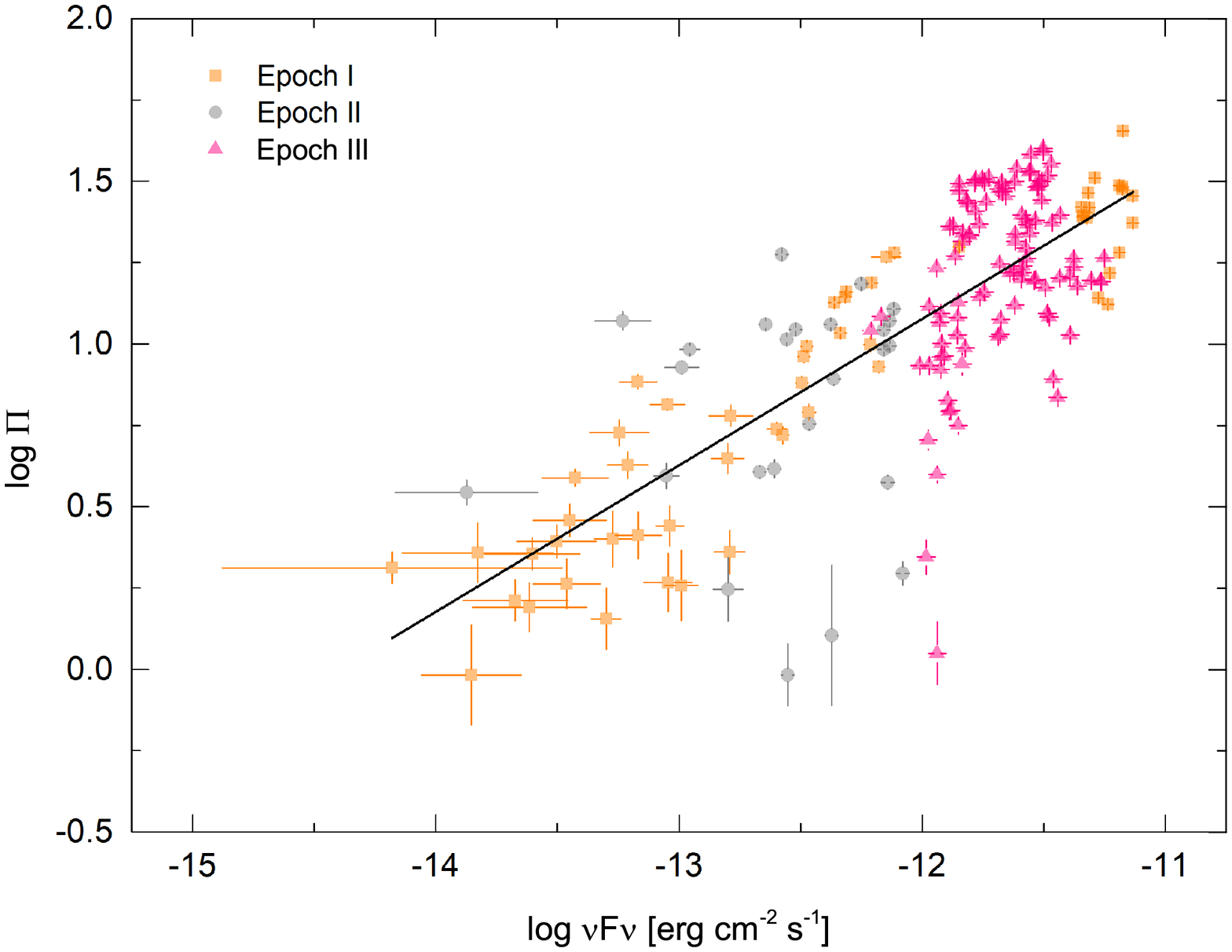}}
    \end{minipage}%
    \caption{Left panel is the plot of $\log E^2 dN / dE$ ($\gamma$-ray) vs. $\log\nu F_{\nu}$ ($V$ band). We subtract a base level flux $3.09\times10^{-13}\rm{erg\,cm^{-2}s^{-1}}$ from the $\nu F_{\nu}$ at $V$ band, because the nonvariable component (including contributions from host galaxy etc.) should not be included in the variation analysis. And the right panel is the plot of $\log \Pi \sim \log \nu F_{\nu}$. The orange squares, gray circles and pink triangles represent the data in Epochs I, II, and III, respectively. }\label{figure10}
\end{figure}

  In Figure \ref{figure10}, it is evident that the logarithm of optical and that of the $\gamma$-ray flux is linearly correlated, the statistic slope is $0.78\pm0.14$ with Pearson's $r=0.71$. The upper limit of the slope approaches 1.
Referring to Table \ref{Tab:index}, if the variation is caused by $N_e$, the predicted slope is 1 for EC and 2 for SSC. Thus, we can conclude that the variation of $N_e$ in the SSC process cannot explain the slope. One can also rule out the case in which the variation of fluxes is due to the change of $B$ in EC process. If the Doppler factor $\delta$ is the main parameter, one needs to discuss the range of $\frac{3+\alpha_{\gamma}}{3+\alpha_o}$ and $\frac{4+2\alpha_{\gamma}}{3+\alpha_o}$. The range of $\alpha_o$  can be obtained from Figure \ref{figure6} via $\alpha_o=(V-R)/2.5\log(\nu_V /\nu_R)$ , i.e., from about 2.18 to 3.27. In Figure \ref{figure9}, $\alpha_{\gamma}$ ranges from 0.834 to 1.566. Thus, the possible ranges of $\frac{3+\alpha_{\gamma}}{3+\alpha_o}$ and $\frac{4+2\alpha_{\gamma}}{3+\alpha_o}$ are $[0.61,0.88]$ and $[0.90,1.38]$ respectively. Therefore, both EC and SSC varying with $\delta$ are possible, since the slope $0.78\pm0.14$ is in these ranges.

\subsection{Polarization and Optical $V$ Band} \label{Sec:pol}
In the upper right panel of Figure \ref{Fig:LCCF}, we obtain a $3\sigma$ correlation between PD and radio light curves. This suggests that variation of PD is correlated with optical flux. So we plot  $\log \prod$ versus $\log \nu F_{\nu}$ in the right panel of Figure \ref{figure10}, where the $\log\prod$ is strongly related to $\log \nu F_{\nu}$ with Pearson's $r=0.77$, and the slope is $0.45\pm0.03$. In \citet{1990apj...350...536C}, the polarization was shown to be a function of $\theta$, where $\theta$ is the angle between the observer's line of sight and the moving direction of the radiative blob. {So we take the empirical relation that $\prod\sim A\sin^n\theta'$, where $n$ is a positive real number, and $\theta'$ is the {viewing angle} in comoving frame of the blob.  \citet{2013mnras...436...1530R} considered $n=2$, based on the study of \citet{Lytikov:2005}, which tells that PD reaches its maximum when $\theta'=\pi/2$, and fall to minimum at $\theta'=0$. In the observer frame, we have}
\begin{equation}\label{eq7}
  \prod\sim\delta^n\sin^n\theta\equiv\delta^{n(1+\alpha_\theta)},
\end{equation}
where $\alpha_\theta=\log_\delta{\sin\theta}$ is the index related to $\delta$.
Thus, $\prod$ is maximal for $\theta\sim\frac{1}{\Gamma}$, corresponding to $\theta'=90^{\circ}$ in the comoving frame \citep{2010ijmpd...19...701N}.  When the blob weaves in our line of sight, $\theta$ will decrease, passing $\frac{1}{\Gamma}$, at which PD achieves its maximum, and then reaches $\theta_{\min}$, at which the flux reaches its maximum. When the blob weaves out, $\theta$ will pass  $\frac{1}{\Gamma}$ again, leading to another peak in the light curve of PD. Thus, one peak of flux is accompanied by two peaks of PD for one outburst for the line of sight inside the beaming cone of the blob.
 However, the sampling of PD is sparse, so that we cannot verify this correspondence from the two light curves directly. A better sampling example, i.e., the intra-day variation study of S5 0715+714, seems to support such correspondence, see Figure 1 in \citet{2015apj...809...130C}. Based on Equation \ref{eq7}, the linear correlation between $\log\prod$ and $\log\nu F_{\nu}$, i.e., $\frac{2(1+\alpha_{\theta})}{3+\alpha_o}$, can be derived in the case that variations of PD and flux are due to the variation of $\delta$, see Table \ref{Tab:index}.
Since $\delta>1$, $\sin\theta<1$, one has $\alpha_{\theta}<0$. Having $n=2$ and $\alpha_o\in [2.18,3.27]$, one has $\frac{2(1+\alpha_{\theta})}{3+\alpha_o}<0.39$.  For $n=3$, one has $\frac{2(1+\alpha_{\theta})}{3+\alpha_o}<0.58$. Thus, the observed slope ($0.45\pm0.03$) can be explained from Equation \ref{eq7}. The index $n$ is model dependent, the correlation analysis here can constrain the polarization model.

The variations of $N_e$ and $B$ cannot lead to the variation of PD in models. The significant correlation suggests that Doppler factor is the key parameter that leads to the variations of fluxes and PD.  Finally, the variation of the Doppler factor is mainly due to the variation of observing angle $\theta$. For this target, it was found that the jet component position angles are nonlinearly distributed, and the jet position angles depend on frequencies at radio bands \citep{2004aa...421...839A}. \citet{Karamanavis2016} studied the dynamical structures based on VLBI images at different radio frequencies, and found that the viewing angle of the inner and outer jet are $3^{\circ}$ and $1^{\circ}$, respectively. They also derived that the jet opening angle is $(3.8\pm0.5)^{\circ}$, and the downstream of jet (away from the core with $1$ mas) bends toward us. Since $\gamma$-ray and optical bands are significantly correlated with radio 15GHz, and they { are located} in the upstream of the radio core, one can expect that radiative blobs of the $\gamma$-ray and optical follow a curved trajectory. Note that the radio light curve has a long-term increasing trend in Epoch III, which does not occur in the $\gamma$-ray and optical light curves. This can be understood if $\gamma$-ray and optical emitting regions have the same viewing angles, which are different from that of the downstream radio emission regions.
The curvature of the jet leads to the different orientation of emitting zones for different frequencies \citep{2017nature...552...374R}. The moving direction of the radiative blob changes when the blob propagates along the jet, which can be realized in the jet precession models \citep{Abraham:2000,Sobacchi:2017}. It is noted that if the trail of emitting regions is helical, the distances from the jet base obtained from time lags would be the upper limit. The method to measure the helical trajectory will be an interesting future work. \par

\subsection{Variations of Color Index and $\gamma$-Ray Spectral Index}{\label{sec:index}}

The color index versus radio case shows a peak at the 2$\sigma$ significance level. Thereby, it is necessary to explore the relationship between the color index and fluxes at different frequencies.  As for the quiescent state (Epoch II), the large range of spectral index and tiny range of flux below 0.25 mJy make it too difficult to determine the trend of the color index variation. The sparse distribution of the spectral index is probably caused by extra emissions from disk, BLR, torus, or the combination of them. \citet{Abdo:2010} derived that the bolometric luminosity of BLR is about $3.7\times 10^{45}$  erg s$^{-1}$, equivalent to 0.03 mJy flux at $V$ band. Considering the Eddington limit of the quasar luminosity \citep{Shapiro:1983}, i.e., $L_{Edd}=1.3\times10^{46} {\rm erg \, s}^{-1} (M/10^8 M_{\bigodot})$, this predicts an extreme 1 mJy flux, if all energy is released at $V$ band. The broadband emission will significantly reduce the flux at $V$ band. Besides, photons from the accretion disk have very small PD, which cannot explain the PD variation at optical band. Due to $z=1.839$, the radiation of dust torus  will mainly shift to far infrared. Thus, contributions from these components can be ignored when the target is in the active state.

The diagram of $V-R$ versus fluxes at $V$ band is shown in Figure \ref{figure6}. Points in Epoch II (the quiescent state) and Epoch III (the active state) are marked with gray and pink color, respectively. The color index in the quiescent state is scatter, while it has a weak RWB trend (with Pearson's $r=0.32$) in the active state. For $F_{\nu}>0.9$ mJy, both a saturation and BWB trend are likely.
In theory, there are several models that can explain the RWB trend. First, \citet{2006aa...453...817V} found an RWB trend for 3C 454.3, and the color index is saturated when fluxes approach the maximum \citep{2006aa...453...817V, 2010pasj...62...645S}. The explanation for the RWB trend is that both accretion disk and jet contribute to the fluxes. The accretion disk contributes a bluer component to the broadband SED, while the synchrotron emission of the jet  contributes a redder component to the continuum. {Secondly}, the RWB phenomenon could also be interpreted by the shock in the jet model \citep{1998aa...333...452K}. When the cooling time scale is roughly the same as the accelerating time scale, the simulation of light curves shows a RWB trend. The BWB trend is due to the fact that the cooling time scale is larger than the accelerating time scale for electrons. However, the accelerating time scale is determined by the strength of the shock.

The twisted jet model is the third model to explain the color index behavior. Suppose the spectrum of radiation is {of} power law $F'_{\nu'}\propto \nu'^{-\alpha_o}$ in the jet comoving frame.  The observed frequency and flux are relativistically boosted via $\nu\propto\delta(\theta)\nu'$ and $F_{\nu}\propto\delta^{3+\alpha_o}(\theta)F'_{\nu'}(\nu)$ \citep{1995pasp...107...803U}. When $\delta$ increases, peak frequency will move from lower frequency to the higher one, the spectral index at a fixed wavelength will undergo variation.  For the observing $V$ and $R$ bands, the amplitude of the Doppler factor is the same for both $V$ and $R$, so  one has $F_{\nu_V}/F_{\nu_R}=(\nu_V/\nu_R)^{-\alpha_o}$. Since  $\nu_V/\nu_R>1$, $F_{\nu_V}/F_{\nu_R}$ is larger or smaller than {unity} for $\alpha_o<0$ or $\alpha_o>0$, respectively.
 Correspondingly, a BWB or RWB trend will occur for $\alpha_o<0$ or $\alpha_o>0$ in the optical bands.

  Since $\alpha_o$ is in the range $[2.18,3.27]$, both $\nu_V$ and $\nu_R$  are higher than peak frequency in SED, which is evident in the broadband SED plot given by \citet{Abdo:2010}.  \citet{2016mnras...462...1508G} showed that the peak frequency is more variable than other frequencies, which can be interpreted in this twisted jet model.  After the peak frequency passes through the observed frequency, the object shows the BWB trend. For PKS 1502+106, there is an RWB trend below $\sim0.85 \,\rm{mJy}$, and a less significant BWB trend beyond that, see Figure \ref{figure6}. Combined with correlation analysis, the curvature effect is a better choice to explain the color index behavior.   Other models cannot be ruled out by the color index analysis alone.

\begin{figure}[ht]
    \begin{minipage}[t]{0.9\linewidth}
        \centerline{\includegraphics[scale=0.35]{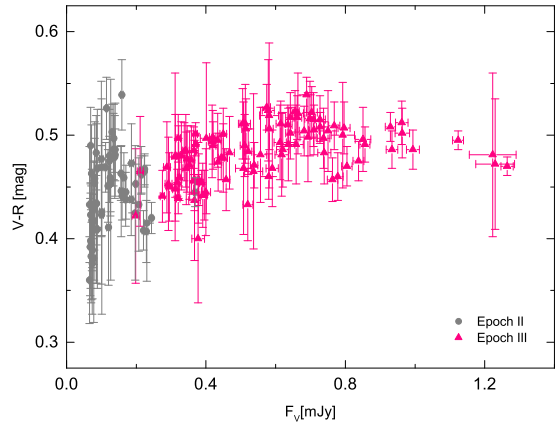}}
    \end{minipage}
    \caption{$V-R$ color index vs. $V$ band flux (in units of mJy) is plotted. The pink triangles and  gray circles represent data in Epoch III and Epoch II, respectively.}\label{figure6}
\end{figure}

The spectral indices $-\alpha_{\gamma}$ of $\gamma$-ray are obtained by linearly fitting $\gamma$-ray fluxes of seven energy grids, while the $\nu F_{\nu}$ $\gamma$-ray fluxes (in units of ${\rm erg \, cm^{-2} \, s^{-1}}$) are obtained by integrating over the $0.1-300$ GeV range.  The  $-\alpha_{\gamma}$ versus  $\log \nu F_{\nu}$ is plotted in Figure \ref{figure9}. The linear fit, with slope $-0.480\pm 0.036$ and  Pearson's $r=-0.816$, indicates that the spectral index is anticorrelated with the flux. This is a softer when brighter (SWB) trend.  Thus, the variations of spectra at optical and $\gamma$-ray are similar, i.e., turning softer when brighter. Meanwhile, the emitting regions of optical and $\gamma$-ray bands are close, which is presented in the previous section. It is likely that the SWB trend is due to the intrinsic property, such as the {evolution of injected particle distribution.}
By studying the continuous equation of injected particles, it was shown that the spectral slope of most energetic particles is steeper than that of the less energetic particles, regardless of whether the radiation is in the fast cooling or slow cooling phases \citep{2002aa...386...833G, 2010mnras...387...1669G}. {  Such intrinsic {spectral} evolution can also explain the RWB trend, but the amplification of fluxes needs more injected electrons. However, if the variation is mainly due to $N_e$,  the slope in Figure \ref{figure10} is predicted to be 1 for EC and 2 for SSC. Another possible reason for the SWB behavior is the curvature effect. {The analysis of flux behavior} also applies to the $\gamma$-ray case, the RWB and SWB trend can both be derived in the modulation of Doppler effects.  \citet{Abdo:2010} (see Figure 11) showed that  the peak frequencies of broadband SED for the low and high bumps are lower than the observing optical and $\gamma$-ray bands, respectively. Based on the SED, both RWB and {SWB} trends are natural results of the curvature effect.
\citet{2004aa...421...839A} and \citet{Karamanavis2016} presented that the  jet  of PKS 1502+106 is twisted.  Thus, the observed variation of $\gamma$-ray spectral index may also be due to the curvature effect. }

\begin{figure}[ht]
        \centerline{\includegraphics[scale=0.35]{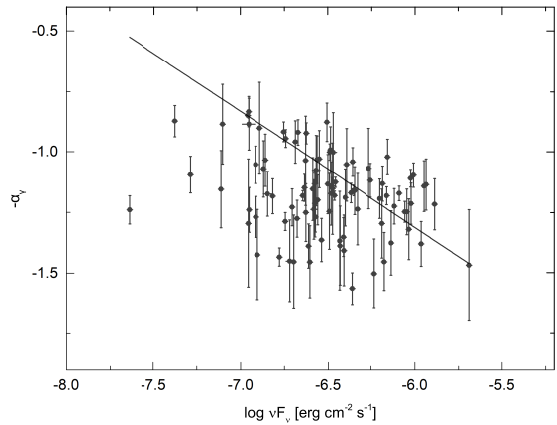}}
         \caption{Distributions of $-\alpha_{\gamma}$ versus $\log \nu F_{\nu}$ of $\gamma$-ray is plotted.
    The {linearly fitted} slope is $-0.480\pm 0.036$ with Pearson's $r=-0.816$.} \label{figure9}
\end{figure}

\section{Conclusion} \label{sec:conclusion}

We {gather the} multifrequency data of PKS 1502+106, including nearly nine years of data of $\gamma$-ray, optical, and radio.
From LCCF calculations, we find that the $\gamma$-ray, optical $V$ band, and PD light curves are correlated with the radio 15 GHz light curve with significance larger than $3\sigma$.
 Based on the FR/RSS MC procedure, the $\gamma$-ray leads the radio with $60_{-10}^{+5}$ days, and leads  $V$ band with $13^{+37}_{-30}$ days. According to LCCFs of both whole period and separated period data, we learn that the sparse sampling is responsible for the plateau which appears in the LCCF of $V$ band versus radio. The distance between $\gamma$-ray and radio core regions is $3.18^{+0.5}_{-0.27}$ pc, which is consistent with the result of \citet{Karamanavis:2016}. The optical and $\gamma$-ray emitting regions are almost the same. Referring to the distance from the $15$ GHz core region to the jet base, the $\gamma$-ray emitting region {is located} less than $1.2$ pc away from the jet base. The possibility of $\gamma$-ray photons produced inside the BLR cannot be ruled out. We find significant linear correlations in both $\log E^2dN/dE\sim\log\nu F_{\nu}$ and $\log\prod\sim\log\nu F_{\nu}$ plots.  Both EC and SSC processes are possible to produce the $\gamma$-ray photons, which agrees with the broad-band SED fitting result \citep{Abdo:2010}. The correlation between PD and $V$ band fluxes can be explained if PD are mainly due to the observing angles.
A less significant RWB trend is found for $V-R$ at the active state, which can be explained by the multiple components model, the shock in the jet model and the twisted jet model.
The spectral index of $\gamma$-ray shows an SWB trend, roughly the same with the RWB trend, which can be explained by the intrinsic spectra evolution of radiative particles and the curvature effect. Based on these findings, the various variation phenomena of PKS 1502+106 can be understood in a unified physical picture, i.e., the radiative blobs trace the curved trajectories, and the variation of viewing angles leads to the variation of the Doppler factor, which further affect the fluxes, PDs, and spectral indices.

\acknowledgments

{We thank the anonymous referee for helpful comments.} The authors are grateful to N. Ding,  Q. Q. Xia, and J. R. Xu for useful discussions.
This work has been funded by the National Natural Science Foundation of China under grant No. U1531105, No. 11403015 and No. 11873035.
Data from the Steward Observatory spectropolarimetric monitoring project were used. This program is supported by Fermi Guest Investigator grants NNX08AW56G, NNX09AU10G, NNX12AO93G, and NNX15AU81G.
This research has made use of data from the OVRO 40 m monitoring program (\cite{Richards:2011}) which is supported in part by NASA grants NNX08AW31G, NNX11A043G, and NNX14AQ89G and NSF grants AST-0808050 and AST-1109911.



\end{document}